\documentclass[preprint]{revtex4}
\usepackage{amssymb,amsmath}

\begin{document}

\draft

\title{ On the Existence of Potential Landscape
  in the Evolution of Complex Systems  }

\author{ Ping Ao$^{\dag}$,
      Chulan Kwon$^{*}$, and Hong Qian$^{ \% }$ }
\address{ $^{\dag}$Department of Mechanical Engineering,
               University of Washington, Seattle, WA 98195, USA \\
          $^{*}$Department of Physics, Myongji University, Namdong San 38-2,
                Yongin, Kyonggi-Do, 449-728, Republic of Korea  \\
          $^{\%}$Department of Applied Mathematics,
               University of Washington, Seattle, WA 98195, USA }
%          $^{\ddag}$Department of Physics,
%               University of Washington, Seattle, WA 98195, USA   }
% {\bf Corresponding author:}
%    $^{\%}$E-mail: aoping@u.washington.edu, Tel: 206-543-7837,
%    Fax: 206-685-8047

\date{Sept. 3, 2005; updated March 20 (2007) for xxx.lanl }

%\noindent
\begin{abstract}
 A recently developed treatment of stochastic processes leads to
 the construction of a potential landscape for the dynamical evolution
 of complex systems.  Since the existence of a potential function in generic settings
 has been frequently questioned in literature, here we study several
 related theoretical issues that lie at core of the construction.
 We show that the novel treatment, via a transformation, is closely related to
 the symplectic structure that is central in many branches of
 theoretical
 physics.  Using this insight, we demonstrate an invariant under the
 transformation.  We further explicitly demonstrate, in one-dimensional case,
 the contradistinction among the new treatment to those of Ito and
 Stratonovich, as well as others.
 Our results strongly suggest that the method from statistical
 physics can be useful in studying stochastic, complex systems in
 general. \\
 {\ }  \\
 To cite published version: \\
  P. Ao, C. Kwon, H. Qian, 2007, On the existence of potential
  landscape in the evolution of complex systems,
  Complexity {\bf 12}: 19-27.
\end{abstract}

%\pacs{PACS numbers: }

\maketitle

\section{ Introduction }

Stochastic differential equations (SDEs) were first introduced in
studying Brownian motions originally discovered in biology
\cite{wax,vankampen}. Other developments in complex systems, from
Darwin's evolutionary theory \cite{ao2005}, the dynamics of gene
regulatory networks \cite{zhu}, the cycle kinetics in biochemical
networks \cite{qian}, to the landscape paradigm
\cite{sherrington}, all have suggested that SDE is a useful and
effective mathematical description for a wide range of dynamical
processes \cite{vankampen,gardiner}. A distinct difference between
the SDEs from biology and those widely studied in classical
physics, usually called Langevin equations, is the absence of the
detailed balance condition in the former. Hence the existence of a
potential landscape with both local and global meanings has been
frequently called into doubt \cite{crutchfield}. During the study
of the robustness and stability of a genetic switch involved in
phage lambda, a bacterium killing virus, one of us has discovered
a mathematical structure embedded in the SDE.  This structure
allows a direct quantification of the global stability of systems
with fluctuations.  Such unique structure leads a transformation
of a SDE to a mathematical form familiar to theoretical
physicists. The transformed SDE has a potential function, a
friction matrix, a transverse force, and a random noise, together
with a relationship between the noise and the friction similar to
the fluctuation-dissipation relation \cite{ao2004,aoyin2006}.
These four dynamical components can be obtained constructively
from a SDE.

The mathematical structure and related transformation have
immediate applications.  With the aid of this method, the
outstanding problem of the stability in phage lambda genetic
switch has been solved \cite{zhu}.  More specifically, the
relative stability of two fixed points in a dynamical system, in
the presence of noise, can be uniquely determined.  The
connections of this mathematical structure to physics, and
detailed behaviors near a fixed point, stable or unstable, have
been analyzed rather exhaustively \cite{kat1}. However, it is also
noticed that in more general situations, the new transformation
may give a result different from the conventional ones from the
traditional integration of a SDE, such as by the Ito integration
or by the Stratonovich integration \cite{ao2004,kat1}. The unique
structure suggests yet another stochastic integration. The purpose
of the present paper is to explore this suggestion and to put it
into a statistical physics perspective. In this way the
application of statistical physics to complex dynamics is
explicitly suggested.

In the following, we first summarize in section II the
transformation from the standard SDE into the transformed
stochastic differential equation. We point out an interesting
invariant of the transformation in section III, suggesting that
the transformation might have a deeper mathematical root. In
section IV we explicitly demonstrate the transformation in a
special one dimensional case, where analytical results have been
known in literature. A comparison to the Ito and Stratonovich
integrations is made. Implications to various realistic
applications are discussed in section V, and we conclude in
section VI. Some mathematical details are provided in the
Appendix.

\section{ Review of the Transformation }
\label{section2}

Let us consider a system described by state vector, $ q^{\tau} =
(q_1, ... , q_n )$, in $n$ dimensional phase space, $\mathbb{R}^n$.
Here the superscript $\tau$ denotes the transpose. The dynamic equation
describing the time evolution of the system is the usual Langevin
like, standard SDE,
\begin{equation}
   \dot{q} = f(q) + \zeta (q,t)~, \label{langevin-1}
\end{equation}
a shorthand writing for ${d q_i}/{d t} = f_{i}(q)+
\zeta_i(q,t)$. Here $\zeta^{\tau}(q,t) = (\zeta_1(q,t), \ldots ,
\zeta_n(q,t))$ denotes stochastic forces, assumed to be
represented by Gaussian white noises, with zero mean, $\langle
\zeta(q,t)\rangle = 0$, and variance
\begin{equation}
   \langle \zeta(q,t) \zeta^{\tau}(q,t') \rangle = 2 D(q)
    \delta (t-t') \; ,
\end{equation}
where $D$ is a positive semi-definite diffusion matrix.
The angular brackets denote the average over noise
distribution, and $\delta(t)$ is the Dirac delta function. The
temperature in the present paper, whenever it can be defined, is
always set to 1. This is equivalent to absorbing the temperature
into the potential function to be discussed below.

It was noticed \cite{ao2004,kat1,aoyin2006} that this equation may
be transformed into the form
\begin{equation}
  [ S(q) + T(q) ] \dot{q} = - \nabla \phi(q) + \xi(q,t) \; ,
\label{gauge}
\end{equation}
where the matrix $S$ is positive semi-definite and symmetric, the
matrix $T$ antisymmetric.  In analogous to the overdamped Brownian
system in physics, here $-S\dot{q}$ plays the role of a
dissipative force with a non-negative energy dissipation rate
$\dot{q}^{\tau}S\dot{q}/2$. The matrix $T$ plays a role similar to
that of a magnetic field. The potential function is $ \phi(q)$.
The new stochastic force $\xi^{\tau}(q,t)=(\xi_1(q,t), ... ,
\xi_n(q,t))$, which has the same origin as $\zeta(q,t)$, is chosen
to satisfy
\begin{equation}
  \langle \xi(q,t) \xi^{\tau}(q,t') \rangle = 2 S(q) \delta(t-t') \; ,
\end{equation}
with zero mean. The steady state distribution for Eq.~(\ref{gauge}),
if exists, would be Boltzmann-Gibbs like
\begin{equation}
   \rho(q,t=\infty) \propto \exp\{ - \phi(q) \} \, .
\label{stationary}
\end{equation}

The transformation from Eqs.~(1) to (3) is rather remarkable in
that the original SDE in Eq.~(\ref{langevin-1}), with no explicit
potential function, is transformed into the dynamic equation in
Eq.~(\ref{gauge}), governed by the potential $\phi$, which can be
obtained without actually solving the time-dependent equation. The
usual assumption of zero probability current or detailed balance
as the potential condition is not required in the passage from
Eq.~(\ref{langevin-1}) to Eq.~(\ref{gauge}). As pointed out,
the results obtained in this way rely on the partial differntial
equation for the probability density, $\rho(q,t)$, that is
different in general from those
based on either Ito or Stratonovich stochastic integration. Thus
several questions naturally arise: Can this difference be more
clearly demonstrated? Are there further implications of this
transformation? In the following we attempt to give an affirmative
demonstration to those questions from two different angles: the
invariant of the transformation and the treatments of SDE's.

\section{ Invariant of the Transformation }

Here we wish to point out an invariant for potential function
during the transformation from Eq.~(1) into (3) within the
framework of Langevin equation, Eq.(6), and the Fokker-Planck
equation (Klein-Kramers equation), Eq.(7), two of most important
equations in the description of nonequilibrium processes in
physics. This invariant suggests a generic nature of the present
transformation deeply rooted in statistical physics.

Let us consider the following nonlinear stochastic dynamics
familiar in physics, an example for noise acting only on half of
state variables:
\begin{equation}
  \left\{  \begin{array}{lll}
     \dot{q} & = & { p \over m }  \\
     \dot{p} & = & - [S(q) + T(q)] {p \over m } - \nabla_q \phi (q)
                 + \zeta(q,t)
            \end{array} \right.
\label{inner}
\end{equation}
Here $q,p$ are $n$-dimensional vectors, $m$ a parameter, $S$ an
$n\times n$  semi-positive definite symmetric matrix, $T$ an
$n\times n$ antisymmetric matrix, and $\phi$ a potential function.
The subscript $q$ in gradient indicates that it operates on $q$
only. There is an additional relationship, due to Einstein, between
the noise $\zeta$ and $S$:
$\langle \zeta(q,t) \zeta^{\tau} (q,t') \rangle = 2
S(q) \delta(t-t') $. This equation is the Kramers' dynamics in
the standard Langevin form in physics \cite{vankampen}.

The probability distribution function $\rho(q,t)$ corresponding to
the dynamical equation satisfies the Fokker-Planck equation
\cite{risken}, or Klein-Kramers equation \cite{vankampen},
\begin{equation}
 \begin{array}{l}
   \left\{ {\partial_t} + {{p}\over{m}}\cdot\partial_q
    + \overline{f}(q,p) \cdot \partial_p  \right.  \\
   {\ } {\ } {\ } \left. - \partial_{p} ^{\tau} S(q)
    \left[{{p}\over{m}} + \partial_p \right]\right\} \rho(q,p,t) = 0 .
 \end{array} \; .
\end{equation}
Here $\overline{f}(q,p) = {p^{\tau} }T(q)/m - \phi_q (q)$ with
$\phi_q (q) = \partial_q \phi(q)$.

A stationary state, while $\partial_t \rho =0$, exists and can
usually be reached after a long time of dynamical evolution. From
the analogy to physics, the final distribution from this dynamics
is given by the Maxwell-Boltzmann-Gibbs distribution:
\begin{equation}
 \rho(q,p,t=\infty) \propto \exp\left\{ - \left[{p^2 \over {2m}}
                   + \phi(q) \right] \right\} \; .
\end{equation}
There is a clear separation of variables $q,p$ in the final
distribution function. This suggests that the mass of the particle
is not an essential quantify to determine the distribution in the
$q$-space, a fact well known in classical statistical physics.
Hence the validity of Eq.~(5) is implied.

We note that Eq.~(6) is in the form of Eq.~(1). To be more
suggestive, we rewrite Eq.~(6) as
\begin{equation}
  \dot{x} =  {\cal F}(x) + {\cal Z}(x,t)
\end{equation}
Here
\[
  x = \left(  \begin{array}{c} q \\ p \end{array} \right)
\]
is a $2n$-dimensional vector,
\[
  {\cal F}(x) = \left(
   \begin{array}{c} p/m \\ - [S(q) + T(q)] {p / m }
- \nabla_q \phi(q) \end{array} \right) \; ,
\]
and
\[
  {\cal Z} = \left(
 \begin{array}{c} 0 \\ \zeta \end{array} \right) \; .
\]
The $2n\times 2n$ diffusion matrix ${\cal D}$ is defined by
\begin{equation}
 \langle {\cal Z}(x,t) {\cal Z}^{\tau}(x,t') \rangle = 2 {\cal D}(x)
   \delta (t-t')\; ,
\end{equation}
and is
\begin{equation}
 {\cal D}(x,t) = \left( \begin{array}{cc}
                           0 & 0 \\ 0 & S(q)
                             \end{array}\right) ~.
\end{equation}

The important question is whether we can obtain the corresponding
form of Eq.~(3):
\begin{equation}
  [ {\cal S}(x) + {\cal T}(x) ] \dot{x} = - \nabla_x \Phi(x)
       + \Xi (x,t)
\end{equation}
with the same prediction on the potential, that is, without any
change in physics prediction. Here ${\cal T}$ is a $2n\times 2n$
antisymmetric matrix  and ${\cal S}$ a $2n\times 2n$ positive
semi-definite symmetric matrix. We also require
\begin{equation}
 \langle \Xi(x,t) \Xi^{\tau}(x,t') \rangle = 2 {\cal S}(x)
   \delta (t-t')\; ,
\end{equation}
with zero mean of $\Xi$.

The answer is affirmative, demonstrating a very suggestive
invariant of the transformation. We give the result here, which
can be verified directly:
\begin{equation}
 {\cal T}(x)  = \left( \begin{array}{cc}
                           T(q) & I \\ -I & 0
                             \end{array}\right)  ~,~
 {\cal S}(x)  = \left( \begin{array}{cc}
                           S(q) & 0 \\ 0 & 0
                             \end{array}\right) ~.
\end{equation}
Here $I$ is the $n\times n$ identity matrix. The corresponding
expression for ${\cal D}+{\cal Q}=({\cal S}+{\cal T})^{-1}$ is
\begin{equation}
 {\cal Q}(x) = \left( \begin{array}{cc}
                           0 & -I \\ I & T(q)
                             \end{array}\right)  ~,~
% {\cal D}(x,t) = \left( \begin{array}{cc}
%                           0 & 0 \\ 0 & S(q)
%                             \end{array}\right) ~.
\end{equation}
and ${\cal D} $ is precisely given by Eq.(11). The potential is
\begin{equation}
  \Phi(x) = {{p^2}\over{2m}} + \phi(q) \; ,
\end{equation}
also precisely the same as that before the transformation. Thus,
once the form of Eqs.~(3) and (6) are reached, further
manipulation of equations according to the transformation does not
change the physical prediction.

\section{ Treatments of Stochastic Differential Equations }

The kinetic momentum in Eq.~(6) can be eliminated, resulting in the
following equation:
\begin{equation}
     m \ddot{q}  =  - [S(q) + T(q)] \dot{q} - \nabla_q \phi(q)
                 + \xi(q,t) \; .
\end{equation}
Apparently, one would be able to take a zero mass limit by setting
the right hand side of Eq.~(17) zero. Carrying out this procedure,
we have, copying Eq.~(3),
\begin{equation}
  [S(q) + T(q) ] \dot{q} = - \nabla_q \phi (q)
                 + \xi(q,t) \; .
\end{equation}
This can be converted into a more suggestive form, the standard
form of SDE, copying Eq.~(1):
\begin{equation}
 \dot{q} = f(q) +  \zeta(q,t) \; .
\end{equation}
Here $ f(q) =  - [S(q) + T(q)]^{-1} \nabla_q \phi (q) $
and $ \zeta(q,t) = [S(q) + T(q)]^{-1} \xi(q,t) $.

Now, in connection of the singular behavior of the zero mass
limit, we are facing a choice on the proper way to carry out the
integration of Eq.~(19). Different stochastic integration method
would give different final distribution, hence different physical
prediction. Here we illustrate this situation by considering
explicitly the one dimensional case, $n=1$. Higher dimensional
case will be considered elsewhere \cite{ya2006}. In one
dimensional case, $n=1$ and $T = 0$, and Eq.~(3) or (19) becomes
\begin{equation}
 \dot{q} = - D(q)\nabla \phi(q) +  \zeta(q,t) \; ,
\end{equation}
and $S(q) D(q) = 1$. We explore its three different
integrations below.

 {\ }

 {\bf I. Present process}.  If we follow a limiting
process to reproduce the final steady state distribution, the
dynamical equation in this limit would be
\begin{equation}
  {\partial_t} \rho(q,t) = [\partial_{q} D(q) \partial_q
        + \partial_q D(q) \phi_q (q) ] \rho(q,t) \; .
\end{equation}
This is in the form of the celebrated Smoluchowski equation
\cite{gardiner-smo}.  Three different derivations from Eq.~(7) to
Eq.~(21) are provided in the Appendix.  Thus, the present
stochastic integration process overlaps with that of the
Smoluchowski process of strong damping limit. The steady state
distribution would be, if exists,
\begin{equation}
  \rho(q,t=\infty) \propto \exp\{-\phi(q) \} \; ,
\label{ours}
\end{equation}
the same as in Eq.~(5).

 {\ }

 {\bf II. Ito process}. The Ito stochastic integration \cite{ito} is a
strict implementation of Markov process and a martingale: There is
no memory effect of previous dynamics and there is no information
from future. The usual differentiation and integration rules would
not be directly applicable here. Fortunately, this case has been
studied in great details. The corresponding Fokker-Planck equation
is, from Eq.~(20) \cite{gardiner-ito}:
\begin{equation}
  {\partial_t} \rho_I(q,t) = [\partial_{q} \partial_q D(q,t)
        + \partial_q D(q,t) \phi_q (q)] \rho_I(q,t) \; .
\end{equation}
The subscript $I$ indicates that Eq.~(23) is in accordance with Ito
process. The steady state distribution would be, if exists, The
steady state distribution would be, if exists,
\begin{equation}
  \rho_{I}(q,t=\infty) \propto \frac{1}{ D(q) }
                         \exp\{-\phi(q) \}  \; .
\end{equation}

 {\ }

 {\bf III. Stratonovich process}.  In the Stratonovich stochastic
integration, the usual differentiation and integration rules apply
\cite{stratonovich}. This implies that the stochastic process such
specified is a Markov process but not a martingale. Again, the
stochastic integration has been carried out, and the corresponding
Fokker-Planck equation is, from Eq.~(20) \cite{gardiner-str}:
\begin{equation}
 \begin{array}{lcl}
  {\partial_t}\rho_S(q,t)& = & [\partial_q {D^{1\over 2}(q,t)}
                           \partial_q {D^{1\over 2}(q,t)} \\
   &  &  + \partial_q D(q,t) \phi_q (q) ] \rho_S(q,t) \; .
 \end{array}
\end{equation}
Similarly, the subscript $S$ indicates that Eq.~(25) is in
accordance with Stratonovich process. The steady state
distribution would be, if exists, the steady state distribution
would be, if exists,
\begin{equation}
  \rho_{S}(q,t=\infty) \propto \frac{1}{ D^{1\over2}(q) }
                          \exp\{-\phi(q) \}  \; .
\end{equation}

{\ }

 The above one-dimensional examples clearly show that the
present treatment of SDE implied in the transformation may be
regarded as another stochastic integration process. It can indeed
produce results different from those of Ito or Stratonovich
process. One way to understand this difference is to view the zero
mass limit as a singular limit in which specific procedure must be
identified in order to get a well-defined result. As explicitly
demonstrated above, there are relative shifts of density peaks
from one process to another. In addition, the positions of peaks
of the distribution function may not coincide with those of zeros
of the potential $\phi$ or the fixed points of $f$. We have
observed such shifts during numerical and analytical study even
when the diffusion matrix $D$ is constant \cite{kat2}, that is,
even when there is no difference among those of Ito and
Stratonovitch, and more general considerations  \cite{wong}.
Similar shifts have also been observed before \cite{zhou,lindner}.
They are associated with the absence of detailed balance: In the
presence of detailed balance condition and with a constant
diffusion matrix, all those stochastic treatments would lead to
the same potential function.
%The above demonstration
%may provide an theoretical explanation for such shifts: Some
%choices of limiting procedure are more sensitive to the detailed
%balance condition.

\section{Discussions}

With various limiting procedures to integrate the SDE of the form
of Eq.~(1), (19) or (20), which one would be right?  The answer is,
not surprisingly, the procedure has to be determined by the real
problem at hand and any of them can be correct in a practical
situation \cite{mortensen,vankampen,gardiner,risken}. We
illustrate this point by examples.

It should emphasized that it has been shown mathematically that in
general, non-white noise can be constructed and its limiting
process can lead to different stochastic integration with
different Wong-Zakai correction \cite{wong}.  We also note, in the
case of 1d, that the Eq.~(\ref{ours}) corresponds to the choice of
evaluating the function at the end of a time interval. Higher
dimensional case would be different \cite{ya2006}. The Ito
process, Eq.(23), is then a choice of initial point, without
peering into future. The Stratonovich process, Eq.(25), is a
choice of middle point.

It is evident that from the physics point of view the first
process appears to be a natural choice: the presence of small mass
limit encountered frequently. It results in the known Smoluchowski
equation \cite{vankampen,gardiner,risken}. Numerous experiments in
physics exist to support this approach. We wish to point out that
the form of Eq.~(6) allows a generalization into more complicated
situations, such as the colored noise, as done in the dissipative
dynamics \cite{zwanzig,leggett}.  This is also the form used in a
recent development of a mesoscopic, open-system nonequilibrium
thermodynamics \cite{qianpre01}.

From an engineering point of view, everything would be thought as
a limit of continuous process, hence Stratonovich process would be
a natural choice. Indeed, analog experiments have been done to
check this situation. The results are in remarkable agreement with
the Stratonovich stochastic integration \cite{mcclintock}.

From a population geneticist's point of view, Ito process would
appear more natural to model the population dynamics. Indeed this
view has been carried out in detail \cite{li,turelli}. We are,
however, not aware of any precise comparison between
experimental/expirical data and the theoretical calculation which
can single out the Ito process as the appropriate description in
this case.

It is worthwhile to point out here that, in connection with the
invariant discussed in section III, in the corresponding
Fokker-Planck equation for Eq.~(6), the Eq.~(7), there is no need
to differentiate all those stochastic processes discussed in
section IV, because of the special algebraic structure expressed
in Eq.~(11) and (14). This adds a further preference to view the
present process as a ``natural'' one.

\section{Conclusion}

In this paper we have given an explicit demonstration of the
difference between the present integration of the stochastic
differential equation and those according to classical methods in
one dimension. While we believe all those integration methods are
in themselves consistent, we do point point that the present
process has a certain advantage, as demonstration by the invariant
of the transformation and by the zero mass limit. Thus, the
existence of potential landscape in complex systems is put on a
firmer theoretical ground.

\bigskip

\noindent {\bf Acknowledgments.} Discussions with R. Mannella, P.
McClintock, D.J. Thouless, D.B. Walton, L. Yin, and X.-M. Zhu are
highly appreciated. This work was supported in part by the USA
National Institutes of Health through Grant number HG002894
(P.A.).

{\ }

\appendix{{\bf Appendix: Three Derivations of Eq.~(21) } }

The particular form of the partial differential equation in
Eq.(21) has not been widely discussed in literature. Since it
plays an important role in our study of the potential landscape,
we give three derivations which are not related to the usual
treatments of stochastic differential equations \cite{wong}. An
extension of Eq.(21) into higher dimensions is discussed in
Ref.\cite{aoyin2006}.

{\ }

 {\bf First derivation}

 Here we give some justifications, both physical and
mathematical, to the derivation of the Eq.~(21) in one dimension.
The derivations for the equations (23) and (25) are well
documented in the literature. Interested readers should consult
textbooks such as that by van Kampen \cite{vankampen}, Gardiner
\cite{gardiner}, or Risken \cite{risken}.

    We note that there is an alternative expression for
Eq.~(\ref{inner})
\begin{equation}
  \left\{  \begin{array}{rll}
     \dot{q} & = & v \\
     m\dot{v} & = & - [S(q,t) + T(q,t)] v  - \nabla_q \phi (q)
                 + \zeta(q,t)
            \end{array} \right.  \; .
\label{outer}
\end{equation}
% which expresses the slow and fast variables more clearly ({\it
% c.f. } Eq.~(17)).
Eqs.~(\ref{inner}) and (\ref{outer}) are for the
inner and outer solutions to the problem with fast and slow time
scales respectively. In one dimensional case, $n=1$ and $T=0$, we
may rewrite equation (7) as
\begin{equation}
 \begin{array}{lcl}
  \partial_t \rho(q,p,t)
    & = & \left( - {{p}\over{m}} \partial_q
    + \phi_q (q)  \partial_p \right. \\
  & & + \left. \partial_{p} S(q)
    \left[{{p}\over{m}} + \partial_p \right]\right) \rho(q,p,t) \; .
  \end{array}
\end{equation}
The right hand side of Eq.~(28) can be regrouped according to the
following manner:
\begin{equation}
 \begin{array}{l}
  \partial_t \rho(q,p,t) \\
  =  \left(
    \partial_p \left[ S(q) \partial_p + S(q){p\over{m}} \right] +
     \partial_p \phi_q (q) - {p\over{m}} \partial_q
      \right) \rho(q,p,t) \\
  =  \left(
    \left[\partial_p - \partial_q {1\over{S(q)}} \right]
    \left[ S(q) \partial_p + S(q){p\over{m}} \right]
     \right. \\
  {\ } {\ } \left.
      + \partial_q {1\over{S(q)}}
      \left[ S(q) \partial_p + S(q){p\over{m}} \right]
      + \partial_p \phi_q (q) - {{p}\over {m}} \partial_q
      \right) \rho(q,p,t) \\
  =  \left(
    \left[\partial_p - \partial_q {1\over{S(q)}} \right]
    \left[ S(q) \partial_p + S(q){p\over{m}} \right]  \right. \\
  {\ }  {\ }  \left.  + \partial_p
      \left[ \partial_q + \phi_q (q) \right]
      \right) \rho(q,p,t) \\
  =  \left(
    \left[\partial_p - \partial_q {1\over{S(q)}} \right]
     \left[ S(q) \partial_p + S(q){p\over{m}} + \phi_q (q)
       + \partial_q \right] \right. \\
 {\ } {\ } + \left. \partial_q
       {1\over{S(q)}} \left[ \partial_q + \phi_q (q)  \right]
      \right)
      \rho(q,p,t) \; . \\
 \end{array}
\end{equation}

    There are two lines of attack for this problem in the
literature.  One was taken by Kramers himself and Chandrasekhar
\cite{wax}, the other is based on the projection operator approach
\cite{gardiner}. The essential assumption in the latter approach
is that the momentum $p$ stationarity is achieved instantaneously
when the mass $m$ is small. It takes the time of order $m/S(q)$
for it to reach equilibrium. Because we are interested in the
dynamics on the time scale $\Delta t \gg m/S(q)$, we may regard
the dynamics of the kinetic momentum $p$ is in an instantaneous
equilibrium with the dynamics of the coordinate $q$. We may then
assume the trial solution for the distribution function in the
following form:
\begin{equation}
  \rho(q,p,t) = \rho(q,t) \rho_K (p,t|q) \; ,
\end{equation}
with conditional probability of $p$ given $q$:
\begin{equation}
   \rho_K (p,t|q) = \sqrt{1\over{2\pi m} }
    \exp\left\{ - {1\over{2m}}
     \left( p - {m \Delta{\phi}_q (q) \over{S(q)}} \right)^2 \right\}
     \; .
\end{equation}
Here $ \Delta{\phi} (q) = - \ln \rho(q,t) - \phi(q) $ and
$\Delta{\phi}_q (q) = \partial_q \Delta{\phi} (q)$. The
distribution of kinetic momentum is Gaussian, in the familiar form
of kinetic energy. At a given moment the average kinetic momentum
is not zero, corresponding to the drifting velocity
$\Delta\phi_q(q)/S(q)$. The drifting velocity is zero at the
eventual equilibrium state, as would expected. We expect that the
error raised from this choice will be in the higher order of
$m/S(q)$ comparing to what would be kept. We will show below that
it is indeed the case.

With the trial solution the contribution of the first term of the
right hand side of Eq.~(29) is:
\begin{equation}
 \begin{array}{lcl}
 K & = &
   \left[\partial_p - \partial_q {1\over{S(q)}}\right]\times \\
   & &  \left[ S(q) \partial_p  + S(q){p\over{m}} + \phi_q (q)
       + \partial_q \right] \rho(q,t) \rho_K (p,t|q)  \\
 & = & \rho(q,t)
   \left[\partial_p - ( - \Delta{\phi}_q (q) -\phi_q(q) + \partial_q )
                       {1\over{S(q)}} \right] \times \\
   & & \left[ S(q) \partial_p + S(q) {p\over{m}} + \phi_q (q)
      \right. \\
   & & \left.   + ( - \Delta{\phi}_q (q) - \phi_q(q) + \partial_q ) \right] \rho_K (p,t|q)  \\
 & = & \rho(q,t)
   \left[\partial_p - ( - \Delta{\phi}_q (q) -\phi_q(q) + \partial_q )
                       {1\over{S(q)}} \right] \times \\
   & & \left[ S(q) \partial_p + S(q) {p\over{m}}
     - \Delta{\phi}_q (q) + \partial_q ) \right] \rho_K (p,t|q)  \\
 & = & \rho(q,t)
   \left[ \partial_p - ( - \Delta{\phi}_q (q)-\phi_q(q) + \partial_q )
                       {1\over{S(q)}} \right] \times \\
   & & \left[ S(q) \partial_p + S(q) {p\over{m}}
         - \Delta{\phi}_q (q) + \partial_q ) \right] \times  \\
   & &  \sqrt{1\over{2\pi m} } \exp\left\{ - {1\over{2m}}
       \left( p - {m \Delta{\phi}_q (q) \over{S(q)}} \right)^2 \right\}
       \; .
    \end{array}
 \end{equation}
Moving $\rho_K$ from the right to left, and grouping all terms, we
have
\begin{equation}
  \begin{array}{lcl}
K & = & \rho(q,t) \times
       \sqrt{1\over{2\pi m} } \exp\left\{ - {1\over{2m}}
      \left( p - {m \Delta{\phi}_q (q) \over{S(q)}}
                \right)^2 \right\} \times \\
   & &   \left[ -{1\over{m}}\left( p-{m \Delta{\phi}_q (q) \over{S(q)}}\right)
         - ( - \Delta{\phi}_q (q) -\phi_q(q)  \right. \\
   & & \left.  - {1\over{m}}\left( p-{m\Delta{\phi}_q(q) \over{S(q)}}\right)
        \partial_q ( {m\Delta{\phi}_q (q)  \over {S(q)} } + \partial_q ))
      {1\over{S(q)}} \right] \times    \\
   & & \left[ - S(q) {1\over{m}}\left( p-{m
         \Delta{\phi}_q (q)\over{S(q)}} \right)
              + S(q) {p\over{m}} - \Delta{\phi}_q (q) \right. \\
   & & \left. - {1\over{m}}\left( p-{m \Delta{\phi}_q (q) \over{S(q)}}\right)
         \partial_q ( {m\Delta{\phi}_q (q) \over{S(q)} } ) \right]  \\
 & = & \rho(q,t) \times
       \sqrt{1\over{2\pi m} } \exp\left\{ - {1\over{2m}}
      \left( p - {m \Delta{\phi}_q (q) \over{S(q)}}
                \right)^2 \right\} \times \\
   & &   \left[ -{1\over{m}}\left( p-{m\Delta{\phi}_q (q)\over{S(q)}}\right)
                 \left( 1 + (\partial_q ({m\Delta{\phi}_q (q)\over {S(q)} }
                 )) \right. \right. \\
   & &   \left. \left.    - \partial_q)
                       {1\over{S(q)}} \right)
      - ( - \Delta{\phi}_q (q) -\phi_q(q) )\right] \times    \\
   & & \left[ - {1\over{m}}\left( p-{m \Delta{\phi}_q (q)\over{S(q)}}\right)
         \partial_q ( {m\Delta{\phi}_q (q) \over{S(q)} } ) \right] \;  .
 \end{array}
\end{equation}

    Performing integration over $p$, all linear terms are zero,
we have
\begin{equation}
 \begin{array}{lcl}
 \int_{-\infty}^{\infty} dp \; K
 & = &
  \rho(q,t) \left( 1 + (\partial_q ({m\Delta{\phi}_q (q)\over {S(q)} } ))
             - \partial_q) {1\over{S(q)}} \right)  \\
  & &  \partial_q ( {m\Delta{\phi}_q (q) \over{S(q)} }) \\
 & = & {m \over S(q) } \rho(q,t)
   \left[S(q) \left( 1 + (\partial_q ({m\Delta{\phi}_q (q)\over {S(q)} }
      )) \right. \right. \\
 & & \left. \left.     - \partial_q) {1\over{S(q)}} \right)
     \partial_q ( {\Delta{\phi}_q (q) \over{S(q)} })
   \right]  \\
 & = & O\left( {m \over S(q) } \right) \rho(q,t)      \\
 & \rightarrow & 0 \;,
  \end{array}
\end{equation}
when ${m \over S(q) } \rightarrow 0$.
Note that $\rho(q,t) = \int_{-\infty}^{\infty} d p \; \rho(q,p,t)
$, integrating Eq.~(29) over $p$, we have
\begin{equation}
 \partial_t \rho(q,t) = \left( \partial_q {1\over{S(q)}}
      \left[ \partial_q + \phi_q (q)  \right]
      \right) \rho(q,t) \; .
\label{theq}
\end{equation}
This is Eq.~(21). The correction is in higher order of $m/S$. This
completes our first derivation of Eq.(21).

{\ }

 {\bf Second derivation}

    We can also follow the original approach of Kramers
for the derivation of Eq.~(\ref{theq}), our second derivation of
Eq.(21). One subtle issue in this approach is realizing that to
obtain $\hat{\rho}(q,t)$ from $\rho(q,p,t)$, one needs to
integrate out $p$ along the curve $q+p/S=$ constant in which
$(p/m)$ is the velocity and $m/S$ is the relaxation time for the
velocity. For any $v$ on the order 1, of course, this correction
is on the order of $m/S$.  For non-constant $S(q)$, the derivation
presented below is in close parallel to that of Kramers, but with
extra mathematical complications.

    We first observe several mathematical relations. \\
 (A.i) From distribution $\rho(x,y)$, the distribution for $z=x+ay$
is $\int \rho(z-ay,y)dy$.  \\
 (A.ii) We have an identity
\begin{equation}
  \begin{array}{l}
   \left(\partial_v-\partial_q\left(\frac{m}{S(q)}\right)\right)\Psi(q,p)
            \Big|_{mv=z-\int S(q)dq}  \\
    = \partial_q{\overline\Psi}\left(q,\frac{z}{m}-
                \frac{1}{m}\int S(q)dq \right)
   \end{array}
\end{equation}
in which ${\overline\Psi}(q,v)=-m \Psi(q,v)/S(x)$.  \\
 (A.iii) Let $v = \frac{1}{m}\left(z-\int S(q)dq\right)$ and its
inverse function be $q=g(z-mv)$.  Then we have, for any $G$
\begin{equation}
 \begin{array}{l}
  \frac{1}{m}\int_{-\infty}^{\infty} dq\ G\left(q,
                \frac{1}{m}\left(z-\int S(q)dq\right)\right) \\
  =  \int_{-\infty}^{+\infty} dv
                \frac{G\left(g(z-mv), v\right)}
        {S\left(g(z-mv)\right)} \\
  \rightarrow \int_{-\infty}^{+\infty}
      \left[dv \frac{G(q,v)}{S(q)} \right]_{q=g(z)}
\end{array}
\end{equation}
when $m\rightarrow 0$.

    Starting with the SDE in Eq.~(\ref{outer}), we have
equation similar to (29):
\begin{equation}
 \begin{array}{l}
  \partial_t \rho(q,v,t) \\
  =  \left(
    \left[\partial_v - \partial_q {m\over{S(q)}} \right]
     \left[{S(q)\over m^2}\partial_v + {vS(q)+\phi_q(q) \over{m}}
       + {1\over m}\partial_q \right] \right. \\
 {\ } {\ } + \left. \partial_q
       {1\over{S(q)}} \left[ \partial_q + \phi_q (q)  \right]
      \right)
      \rho(q,v,t) \; . \\
 \end{array}
\end{equation}
Define random variable $z = mv+\int S(q)dq$ and we have its
distribution according to the relation (A.i),
\begin{equation}
        \rho(z,t) = \frac{1}{m}\int_{-\infty}^{\infty}
                \rho\left(q,\frac{z}{m}
                        -\frac{1}{m}\int S(q)dq\right)dq.
\end{equation}
Using the above relations (A.ii) and (A.iii),
\begin{equation}
 \begin{array}{l}
%\begin{eqnarray*}
   \partial_t \rho(z,t)   \\
  =  \frac{1}{m}\int_{-\infty}^{+\infty} dq \times  \\
 {\ } {\ }  \left[ \partial_q \left({1\over S(q)}\partial_q
        + {\phi_q(q) \over S(q)} \right) \rho(q,v,t)
                \right]_{v=\frac{1}{m}\left(z-\int S(q)dq\right)}  \\
  \rightarrow
        \int_{-\infty}^{+\infty} \frac{dv}{S(q)}
                \left[ \partial_q
        \left({1\over S(q)}\partial_q
        + {\phi_q(q)\over S(q)} \right) \rho(q,v,t) \right]_{q=g(z)} \\
  =  \frac{1}{S(q)}
                \left[\partial_q
        \left({1\over S(q)}\partial_q
        + {\phi_q(q)\over S(q)}\right)S(q)\rho(z,t)\right]_{q=g(z)} \; .
%\end{eqnarray*}
    \end{array}
\end{equation}
Finally, we note that if we define random variable $\zeta=g(u)$,
then the distribution for $\zeta$: $\rho_{\zeta}(y)$ $=$
$S(y)\rho_z\left(\int S(y)dy\right)$. In the limit of
$m\rightarrow 0$, $\zeta=$ $g\left(mv+\int S(q)dq\right)$
$\rightarrow q$, and the distribution for $\zeta$ satisfies the
equation:
\begin{equation}
  \partial_t \rho(q,t) = \partial_q
        \left({1\over S(q)}\partial_q
        + {\phi_q(q)\over S(q)}\right)\rho(q,t).
\end{equation}
Again, this is Eq.~(21).

{\ }

 {\bf Third derivation}

It may appear that the first derivation is too intuitive and the
second one is too formal. One may wonder whether or not there is a
more programmatic derivation. The answer is positive. Here we give
such one where the higher order correction in the small mass limit
would be systematically obtained.

The trial distribution function is assumed in Eqs. (30), (31). We
show it is indeed the right solution from the Klein-Kramers
equation in Eq. (28).

First we write
\begin{eqnarray}
 \rho(q,p,t) & = & \exp \Big( - \Delta \phi(q) - \phi(q)
   - {p^2\over 2m}   \nonumber \\
   & & + A(q,p,t) + m B(q,p,t) \Big)
\end{eqnarray}
The higher order terms in $m$ are included in $B(q,p,t)$. Then Eq.
(28) gives
\begin{eqnarray}
 \lefteqn{-\Delta\phi_t-\phi_t+A_t+mB_t}  \nonumber\\
  & = & {p\over m}(\Delta\phi_q-A_q)-pB_q
   + \phi_q A_p + m \phi_q B_p  + S(q)         \nonumber\\
  &   &  \left(-{p\over m} A_p - p B_p
   + A_p^2 + 2m A_p B_p + m^2 B_p^2 \right)
\end{eqnarray}
where the subscripts denote the derivatives with respect to $t,
q,p$. Then the term of ${\cal O}(1/m)$ yields
\begin{equation}
 \Delta\phi_q - A_q - S(q)A_p = 0~.    \label{A}
\end{equation}
Then we can find $A$ for small $p$, noting $p$ is order of $m$.
\begin{equation}
 A(q,p) = {\Delta\phi_q \over S(q)}p + {\cal O}(p^2)
\end{equation}
This satisfies Eq. (\ref{A}) in the above, where $A_q$ gives
${\cal O}(p)$. This justifies the trial form of the distribution
function used in our first derivation. The rigorous solution for
$A(q,p)$ can be found:
\begin{equation}
 A(q,p) = \left[ 1 - \exp\left(-{p \over
   S(q)} \partial_q\right) \right] \Delta \phi
\end{equation}
We will take only the first order in $p$. Nevertheless, we point
out that this procedure can be carried out to higher orders by
matching the corresponding orders in Eq. (43).

The expression corresponding to Eq. (33) should read as follows:
\begin{eqnarray}
 K & = &\rho(q,t)\times\sqrt{1\over 2\pi m}
    \exp\left\{ -{1\over 2m} \left(p-{m\Delta\phi_q(q)\over S(q)}
    \right)^2\right\}        \nonumber \\
   & & \times\left[-{1\over m}\left(p-{m\Delta\phi_q(q)\over
    S(q)}\right)+\partial_p -\Big\{-\Delta\phi_q(q)-\phi_q(q)
      \right.                \nonumber\\
   & & \left.\left.+{1\over m}\left(p-{m\Delta\phi_q(q)\over
    S(q)}\right)\partial_q\left({m\Delta\phi_q(q)\over S(q)}
        \right)+\partial_q\right\}{1\over S(q)}\right] \nonumber\\
   & & \times\left[ -{S(q)\over m}\left(p-{m\Delta\phi_q(q)\over
    S(q)}\right)+S(q){p \over m}\right. \nonumber\\
   & & \left. -\Delta\phi_q(q)+{1\over m}\left(p-{m\Delta\phi_q(q)\over
    S(q)}\right)\partial_q\left({m\Delta\phi_q(q)\over S(q)}
      \right)\right]                              \nonumber\\
   & = & \rho(q,t)\times\sqrt{1\over 2\pi m}\exp\left\{ -{1\over
    2m} \left(p-{m\Delta\phi_q(q)\over S(q)}\right)^2\right\} \nonumber\\
   & & \times\left[ \left\{ -{1\over m} \left(p-{m\Delta\phi_q(q)\over
    S(q)}\right) \left(1+{1\over S(q)} \partial_q\left( {m\Delta\phi_q\over
    S(q)}\right) \right) \right. \right.  \nonumber \\
   & &  + \left. ( \Delta\phi_q + \phi_q-\partial_q)
      {1\over S(q)}\right\}   \nonumber \\
   & & \left.\times\left\{{1\over m}\left(p-{m\Delta\phi_q(q)\over
    S(q)}\right)\partial_q\left({m\Delta\phi_q\over
    S(q)}\right)\right\} \right. \nonumber \\
   & &  \left. +\partial_q\left({\Delta\phi_q\over S(q)}\right)\right]~.
\end{eqnarray}
According to this, Eq. (34) leads to:
\begin{eqnarray}
 \int_{-\infty}^{\infty}dp K
  & = & \rho(q,t)\left[ -{1\over m}\left\{1+{1\over S(q)}
   \partial_q\left({m\Delta \phi_q\over
   S(q)}\right)\right\} \right. \nonumber \\
  & & \partial_q \left({m\Delta \phi_q\over
   S(q)}\right)
    + {m\over S(q)}\left\{\partial_q\left({\Delta \phi_q\over
   S(q)}\right)\right\}^2                  \nonumber\\
  & & \left. + \partial_q\left({\Delta \phi_q\over
   S(q)}\right) \right]\nonumber\\
  & = & 0~.
\end{eqnarray}
It is exactly zero, not approximate zero as in Eq. (34) Again, Eq.
(21) follows in the zero mass limit.

Above demonstration evidently indicates that the novel stochastic
integration procedure discussed in the present paper has a strong
connection to the overdamping dynamics, the Smoluchowski limit.
They are the same in one dimension, but differ generally in higher
dimensional cases which would be addressed elsewhere.

{\ }

\end{document}